\documentclass[aps,pre,preprint,showpacs,eqsecnum,amssymb]{revtex4}
\usepackage{graphicx}
\begin{document}
%\date{\today}
\title{Heat fluctuations in equilibrium}
\author{Hans C. Fogedby}
\email{fogedby@phys.au.dk}
\affiliation{Department of Physics and
Astronomy, University of
Aarhus, Ny Munkegade\\
8000 Aarhus C, Denmark\\}
\begin{abstract}
The characteristic function for heat fluctuations in a non equilibrium system is
characterised by a large deviation function whose symmetry gives rise to a
fluctuation theorem. In equilibrium the large deviation function vanishes and
the heat fluctuations are bounded. Here we consider the characteristic function for 
heat fluctuations in equilibrium, constituting a sub-leading correction to the large 
deviation behaviour. Modelling the system by an oscillator coupled to an explicit 
multi-oscillator heat reservoir we evaluate the characteristic function.
\end{abstract}
.

\maketitle
%%%%%%%%%%%%%%%%%%%%%%%%%%%%%%%%%%%%%%%%%%%%%%%%%%%%%%%%%
%%%%%%%%%%%%%%%%%%%%%%%%%%%%%%%%%%%%%%%%%%%%%%%%%%%%%%%%%
\section{\label{Intro}Introduction}
%%%%%%%%%%%%%%%%%%%%%%%%%%%%%%%%%%%%%%%%%%%%%%%%%%%%%%%%%
%%%%%%%%%%%%%%%%%%%%%%%%%%%%%%%%%%%%%%%%%%%%%%%%%%%%%%%%%
There is a current interest in the thermodynamics and statistical mechanics of small fluctuating systems in contact with 
heat reservoirs and driven by external forces. This interest stems from the recent possibility of the direct manipulation 
of nano systems and biomolecules in non equilibrium scenarios 
\cite{Trepagnier04,Collin05,Seifert06a,Seifert06b,Imparato07,Ciliberto06,Ciliberto07, Ciliberto08}.
together with the advent of the so-called fluctuation theorems which impose symmetry relations on the non 
equilibrium heat and work probability distributions \cite{Jarzynski97,Kurchan98,Gallavotti96,Crooks99,Crooks00,Seifert05a,Seifert05b,
Evans93,Evans94,Gallavotti95,Lebowitz99,Gaspard04,Imparato06,
vanZon03,vanZon04,vanZon03a,vanZon04a,Seifert05c}. In recent years there has also been increased interest
in the above non equilibrium issues for open quantum systems, see e.g. \cite{Salazar19,Denzler18}.

As an illustrative example we consider a single oscillator coupled to two heat reservoirs at temperatures 
$T_1$ and $T_2$ and characterised by the damping constants $\Gamma_1$ and $\Gamma_2$, respectively \cite{Derrida05,Fogedby11a}.  The representative Langevin equations for the position $x$, 
the momentum $p$, force constant $\kappa$, heat flux $dQ/dt$ from reservoir 1, and noises $\xi_1$ and $\xi_2$ 
are then given by (setting the mass $m=1$)
\begin{eqnarray}
&&\frac{dx}{dt}=p,
\label{in1}
\\
&&\frac{dp}{dt}=-(\Gamma_1+\Gamma_2)p-\kappa x+\xi_1+\xi_2,
\label{in2}
\\
&&\frac{dQ}{dt}=-\Gamma_1p^2+p\xi_1,
\label{in3}
\\
&&\langle\xi_1\xi_1\rangle(t)=2\Gamma_1T_1\delta(t),
\label{in4}
\\
&&\langle\xi_2\xi_2\rangle(t)=2\Gamma_2T_2\delta(t).
\label{in5}
\end{eqnarray}
The characteristic function describing the long time behaviour of the heat fluctuations is given by
\begin{eqnarray}
\tilde C(k,t)=\langle\exp(kQ(t))\rangle=C(k)\exp(t\mu(k)),
\end{eqnarray}
where $\mu(k)$ is the large deviation function. The general Gallavotti-Cohen fluctuation theorem applying
to nonequilibrium systems implies the fundamental symmetry \cite{Gallavotti95,Fogedby11a}
\begin{eqnarray}
\mu(k)=\mu(1/T_1-1/T_2-k).
\end{eqnarray}
For the above example we have explicitly \cite{Derrida05,Fogedby11a}
\begin{eqnarray}
&&\mu(k)=\bigg[\Gamma_1+\Gamma_2-\sqrt{(\Gamma_1+\Gamma_2)^2+4\Gamma_1\Gamma_2f(k)}\bigg],
\\
&&f(k)=T_1T_2k(1/T_1-1/T_2-k),
\end{eqnarray}
where the form of $f(k)$ implies the fluctuation theorem. Disconnecting for example reservoir 1 by setting 
$\Gamma_1=0$ the large deviation function $\mu(k)$ vanishes. The resulting system with an oscillator interacting with a single reservoir is an equilibrium system with bounded heat fluctuations and $\tilde C(k)=C(k)$. 

The purpose of the present paper is to investigate further  the case of fluctuations in equilibrium described by
the characteristic function $C(k)$.  In the case of a thermodynamic variable $x$ the understanding is well known 
and follows from the Boltzmann-Gibbs scheme \cite{Landau80a,Reichl98}.  
The probability distribution $P(x)$ is determined by the entropy $S(x)$ according to $P(x)\propto\exp(S(x))$  \cite{Landau80a,Reichl98}. Expanding $S(x)$ about the its maximum value corresponding to equilibrium,
 i.e., $S(x)=S(x_0)-\text{const.}(x-x_0)^2$, we arrive at the Gaussian distribution $P(x)\propto\exp(-x^2/2\langle x^2\rangle)$, where $\langle x^2\rangle$ is the mean square fluctuation.

However, in the case of the fluctuating heat $Q$ exchanged between a small system and a single heat reservoir
at inverse temperature $\beta=1/T$ (we have set the Boltzmann constant $k_{B}=1$), the heat distribution 
$P(Q)$, surprisingly, does not take a Gaussian form. As discussed in previous papers \cite{Imparato07,Fogedby09a} addressing an over damped oscillator, the distribution has the form $P(Q)=(\beta/\pi)K_0(\beta|Q|)$, where $K_0$ is a 
Bessel function \cite{Lebedev72,Gradshteyn65}. This distribution only depends on the temperature of the 
reservoir, exhibits exponential Boltzmann tails $\sim\exp(-\beta |Q|)$ and diverges logarithmically at small $Q$ as 
$P(Q)\sim -\ln(|Q|)$. 

In general the characteristic function, defined according to \cite{Reichl98}
\begin{eqnarray}
C(k)=\int dQ P(Q)\exp(kQ),
\label{char11}
\end{eqnarray}
is given by the expression
\begin{eqnarray}
C(k)=\frac{Z(\beta+k)Z(\beta-k)}{Z(\beta)^2},
\label{char1}
\end{eqnarray}
where $Z(\beta)$ is the partition function for the system \cite{Landau80a,Reichl98}. We note that in the case 
of a single over damped oscillator with one degree of freedom $Z(\beta)\propto 1/\sqrt\beta$,
i.e., $C(k)=\beta/\sqrt{\beta^2-k^2}$. Likewise,  for a damped harmonic oscillator with two degrees of 
freedom coupled to a reservoir $Z(\beta)\propto 1/\beta$, yielding the characteristic function 
$C(k)=\beta^2/(\beta^2-k^2)$. The corresponding heat distribution is $P(Q)=(\beta/2)\exp(-\beta|Q|)$,
decaying exponentially and exhibiting a cusp at $Q=0$.

We note that in the presence of two temperature-biased heat reservoirs driving the system into a 
non equilibrium state, a finite fluctuating heat flux $q=dQ/dt$ will be established. As a result the integrated
heat $Q(t)=\int^t_0dt'dq/dt'$ will on average grow linearly in time, i.e., $Q(t)\sim qt$. More precisely, 
the characteristic function associated with the probability distribution $P(Q,t)$,
$\langle\exp(kQ(t))\rangle\sim\exp(t\mu(k))$, where $\mu(k)$ is the large deviation function; 
in equilibrium we have $\mu=0$. In that sense the equilibrium heat distribution constitutes the sub-leading 
correction to the large deviation result.

In the present paper we extend the analysis in \cite{Fogedby09a} and consider the case of an explicitly defined
heat reservoir in evaluating the equilibrium heat distribution.  For convenience  we consider the case of a single 
oscillator coupled to single heat bath. The explicit characterisation of the heat bath in terms of a collection of 
oscillators is well known and a prerequisite for a quantum treatment. It is also well known that the so-called
ohmic approximation is equivalent to a standard Langevin/Fokker Planck description. However, we believe
that the present calculation carried out within the multi-oscillator scheme is novel.

The paper is organised in the following manner. In Sec. \ref{heatdef} we define the heat exchanged between the system 
and the reservoir. In Sec. \ref{dis} we discuss a heuristic derivation of the characteristic function, 
yielding Eq. (\ref{char1}). In Sec. \ref{mul1} we discuss the explicit characterisation of the heat bath in terms 
of a collection of oscillators. In Sec. \ref{mulheat} we turn to an evaluation of the heat characteristic function 
in the case of a multi-oscillator heat bath. Since the methods we employ are well known we defer technical details to 
appendix A and B sections.
%%%%%%%%%%%%%%%%%%%%%%%%%%%%%%%%%%%%%%%%%%%%%%%%%%%%%%%%%
%%%%%%%%%%%%%%%%%%%%%%%%%%%%%%%%%%%%%%%%%%%%%%%%%%%%%%%%%
\section{\label{heatdef}Heat}
%%%%%%%%%%%%%%%%%%%%%%%%%%%%%%%%%%%%%%%%%%%%%%%%%%%%%%%%%
%%%%%%%%%%%%%%%%%%%%%%%%%%%%%%%%%%%%%%%%%%%%%%%%%%%%%%%%%
A small system coupled to a heat reservoir constitutes a closed system. Correspondingly, the total energy of system
and reservoir is conserved. However, the small system itself exchanges energy with the reservoir and is 
in this respect and open system. 

Let us characterise the small fluctuating system coupled to the heat reservoir by the fluctuating Hamiltonian 
$H_0(t)=H_0(\{x_n(t)\})$, where $\{x_n\}$ are the degrees of freedom. The time dependence of $H_0(t)$ is due to the heat reservoir and not to an 
applied external time dependent force, i.e, we are not applying an external protocol as is common 
in the context of fluctuation theorems \cite{Jarzynski97,Kurchan98,Gallavotti96,Crooks99,Crooks00,Seifert05a,Seifert05b,Evans93,Evans94,Gallavotti95,
Lebowitz99,Gaspard04,Imparato06,vanZon03,vanZon04,vanZon03a,vanZon04a,Seifert05c}.

The fluctuating heat flux $q(t)$ from the reservoir to the small system is thus given by $q(t)=dH_0/dt$. 
Note that since we are in equilibrium the mean value $\langle q(t)\rangle=0$. Consequently, the
heat $Q(t)$ transmitted in a time span $t$ is $Q(t)=\int_0^t d\tau q(\tau)$, i.e.,
\begin{eqnarray}
Q(t)=H_0(t)-H_0(0).
\label{heat1}
\end{eqnarray}
The transmitted heat $Q(t)$ is a fluctuating quantity and the issue is to determine its stationary probability distribution
$P(Q)$, or, equivalently, its characteristic function
\begin{eqnarray}
C(k)=\langle\exp(kQ(t))\rangle=\int dQ P(Q)\exp(kQ).
\label{char2}
\end{eqnarray}
%
%%%%%%%%%%%%%%%%%%%%%%%%%%%%%%%%%%%%%%%%%%%%%%%%%%%%%%%%%
%%%%%%%%%%%%%%%%%%%%%%%%%%%%%%%%%%%%%%%%%%%%%%%%%%%%%%%%%
\section{\label{dis}Heat distribution - heuristic derivation}
%%%%%%%%%%%%%%%%%%%%%%%%%%%%%%%%%%%%%%%%%%%%%%%%%%%%%%%%%
%%%%%%%%%%%%%%%%%%%%%%%%%%%%%%%%%%%%%%%%%%%%%%%%%%%%%%%%%
Here we present a heuristic derivation of the heat distribution or characteristic function; this  approach was also 
discussed in \cite{Fogedby09a} and is based on the definition in Eq. (\ref{heat1}). From Eqs. (\ref{heat1}) and 
(\ref{char2}) we have
\begin{eqnarray}
C(k)=\langle\exp(k(H_0(t)-H_0(0)))\rangle.
\label{char3}
\end{eqnarray}
Assuming that the system is in equilibrium at time $t=0$ and at time $t$ and, moreover, assuming that $t$ is larger
than the characteristic decay time of fluctuations, the energy fluctuations at time $t=0$ and time $t$ can be assumed
to be uncorrelated and we infer $C(k)\sim \langle\exp(kH_0(t))\rangle\langle\exp(-kH_0(0))\rangle$. An interesting
side issue is the role of finite time correlations; this will, however, not be considered here.
Finally, averaging with respect to the canonical ensemble, $\exp(-\beta H_0)/Z(\beta)$, for a heat reservoir maintained at temperature $T=1/\beta$, we arrive at Eq. (\ref{char1}), i.e.,
\begin{eqnarray}
C(k)=\frac{Z(\beta+k)Z(\beta-k)}{Z(\beta)^2}.
\label{char4}
\end{eqnarray}
In the case of two degrees of freedom, e.g., for a harmonic oscillator coupled to a heat bath, 
where $Z(\beta)\propto1/\beta$  \cite{Landau80a,Reichl98}, we obtain
\begin{eqnarray}
C(k)=\frac{\beta^2}{\beta^2-k^2}.
\label{char5}
\end{eqnarray}
%
%%%%%%%%%%%%%%%%%%%%%%%%%%%%%%%%%%%%%%%%%%%%%%%%%%%%%%%%%
%%%%%%%%%%%%%%%%%%%%%%%%%%%%%%%%%%%%%%%%%%%%%%%%%%%%%%%%%
\section{\label{mul1}Multi-oscillator heat bath}
%%%%%%%%%%%%%%%%%%%%%%%%%%%%%%%%%%%%%%%%%%%%%%%%%%%%%%%%%
%%%%%%%%%%%%%%%%%%%%%%%%%%%%%%%%%%%%%%%%%%%%%%%%%%%%%%%%%
Here we derive the characteristic function introducing an explicit representation of the heat reservoir in terms of 
a system of non-interacting oscillators. In a quantum context this is a standard approach \cite{Ford65,Ford87,
Ford88,Caldeira83a,Caldeira83b}  and we therefore defer technical details to  Appendix \ref{app1}.
The total system consisting of the oscillator coupled to the heat bath is isolated and globally energy conserving.
The oscillator itself can exchange energy with the reservoir and is in this regard an open system. The total
system is described by the oscillator Hamiltonian $H_0$, a Hamiltonian $H_1$ for the heat
bath together with an interaction term $V$ to be specified later. We have
\begin{eqnarray}
&&H_0(x,p)=\frac{p^2}{2m}+\frac{1}{2}m\omega_0^2 x^2,
\label{ham1}
\\
&&H_1(\{x_k,p_k\})=\sum_k\bigg(\frac{p_k^2}{2M}+\frac{1}{2}M\Omega_k^2x_k^2\bigg),
\label{ham2}
\end{eqnarray}
where $m$ is the mass and $\omega_0$ the frequency of the oscillator. The bath oscillators
have mass $M$ and frequencies $\Omega_k$, where $k$ is the wave number.

The analysis is simplified by introducing the complex amplitude variables $a$ and $b_k$, see Appendix \ref{app1},
 yielding the Hamiltonians
\begin{eqnarray}
&&H_0(a)=\omega_0|a|^2,
\label{ham3}
\\
&&H_1(\{b_k\})=\sum_k\Omega_k|b_k|^2.
\label{ham4}
\end{eqnarray}
For the interaction between the oscillator and the heat bath we choose the linear coupling
\begin{eqnarray}
V=\sum_k\lambda_k(ab^\ast_k+a^\ast b_k),
\label{ham5}
\end{eqnarray}
where the coupling strength $\lambda_k$ is assumed weak. By appropriate choice of the phases of $a$ and 
$b_k$ we can ensure that $\lambda_k$ is real. We note that the interaction Hamiltonian $V$ differs
from the shift interaction $H_1=\sum_k\bigg(p_k^2/2M+(1/2)M\Omega_k^2(x_k-x)^2\bigg)$
used by Ford  \cite{Ford65,Ford87,Ford88}. In the present case the interaction $V$ corresponds to the 
rotating wave approximation employed in quantum optics \cite{Gardiner97,Fogedby93,Glauber84}.

In the ohmic approximation \cite{Ford65,Ford87,
Ford88,Caldeira83a,Caldeira83b}  and at long times the amplitude of the oscillator is given by
\begin{eqnarray}
a(t)=\sum_k\frac{\lambda_ke^{-i\Omega_k t}}{\Omega_k-\tilde\omega_0+i\Gamma}b_k,
\label{amp}
\end{eqnarray}
where the renormalised frequency $\tilde\omega_0=\omega_0+\Delta$, the shift $\Delta$ and the
damping $\Gamma$ are given by
\begin{eqnarray}
&&\Delta=P\sum_k\frac{\lambda_k^2}{\omega_0-\Omega_k},
\label{shift}
\\
&&\Gamma=\pi\sum_k\lambda_k^2\delta(\omega_0-\Omega_k).
\label{damp}
\end{eqnarray}
Here $b_k=b_k(0)$ is the initial value of the amplitude of the k-th reservoir mode.
In the limit of a large reservoir time smoothing is implemented and we have 
$\sum_k\lambda_k^2\cdots=\int d\Omega g(\Omega)\cdots$, where $\lambda_k^2$ is incorporated
in the definition of the density of states $g(\Omega)$. For $\Delta$ and $\Gamma$ we then obtain
\begin{eqnarray}
&&\Delta=P\int d\Omega\frac{g(\Omega)}{\omega_0-\Omega},
\label{shift2}
\\
&&\Gamma=\pi\int d\Omega g(\Omega)\delta(\omega_0-\Omega)=\pi g(\omega_0).
\label{damp2}
\end{eqnarray}
It is instructive to show that the oscillator locks onto the reservoir temperature at long times. Inserting the
solution in Eq. (\ref{amp}) in Eq. (\ref{ham3}) and using equipartition of the k-the mode, i.e.,
$\Omega_k\langle|b_k|^2\rangle=T$, we obtain for the mean value of $H_0$
\begin{eqnarray}
\langle H_0\rangle=
\omega_0\sum_k\frac{T}{\Omega_k}\frac{\lambda_k^2}{(\Omega_k-\tilde\omega_0)^2+\Gamma^2}
\sim\omega_0T\int d\Omega\frac{g(\Omega)}{\Omega((\Omega-\tilde\omega_0)^2+\Gamma^2)}\sim T,
\label{mean}
\end{eqnarray}
demonstrating equipartition for the oscillator due to coupling to the heat reservoir
%%%%%%%%%%%%%%%%%%%%%%%%%%%%%%%%%%%%%%%%%%%%%%%%%%%%%%%%%
%%%%%%%%%%%%%%%%%%%%%%%%%%%%%%%%%%%%%%%%%%%%%%%%%%%%%%%%%
\section{\label{mulheat}Heat distribution}
%%%%%%%%%%%%%%%%%%%%%%%%%%%%%%%%%%%%%%%%%%%%%%%%%%%%%%%%%
%%%%%%%%%%%%%%%%%%%%%%%%%%%%%%%%%%%%%%%%%%%%%%%%%%%%%%%%%
Regarding the characteristic function for the heat we obtain, inserting $Q(t)=H_0(t)-H_0(0)$
from Eq. (\ref{heat1}) and averaging over the initial reservoir states $b_k$ according to 
$H_1(0)=\sum_k\Omega_k|b_k(0)|^2$, the functional integral
\begin{eqnarray}
C(k)=\int\prod_k|db_k|^2e^{-\beta H_1(0)}e^{k(H_0(t)-H_0(0))}\bigg/
\int\prod_k|db_k|^2e^{-\beta H_1(0)},
\label{char6}
\end{eqnarray}
where we have used $\prod_k dp_kdx_k\propto\prod_k db^\ast_kdb_k$. Inserting the solution $a(t)$ in 
Eq. (\ref{amp}) in $H_0(\{a\})$ in Eq. (\ref{ham3}) the functional integral (\ref{char6}) has a Gaussian form 
and can be performed using standard techniques \cite{Zinn-Justin89}. By mean of the identity 
$\int\prod_k|db_k|^2\exp(-\sum_{kp}b^\ast_kA_{kp}b_p)\propto(\det A)^{-1}$ the evaluation of (\ref{char6}) 
is reduced to an eigenvalue problem. Deferring details to Appendix \ref{app2} we obtain the characteristic 
function in Eq. (\ref{char5}), i.e., $C(k)=\beta^2/(\beta^2-k^2)$.
%%%%%%%%%%%%%%%%%%%%%%%%%%%%%%%%%%%%%%%%%%%%%%%%%%%%%%%%%
%%%%%%%%%%%%%%%%%%%%%%%%%%%%%%%%%%%%%%%%%%%%%%%%%%%%%%%%%
\section{\label{disc}Discussion}
\begin{figure}
\begin{center}
\includegraphics[width=1.0\hsize]{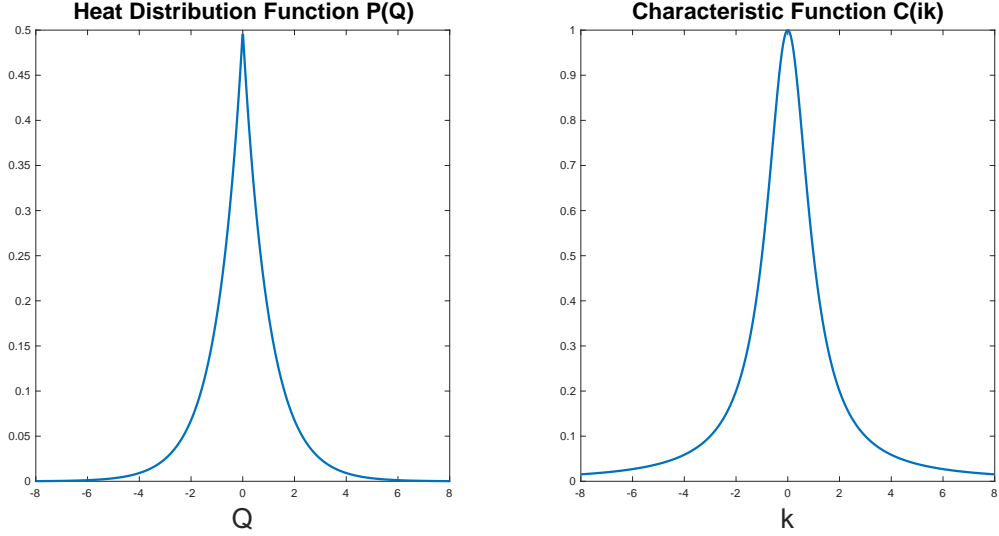}
\end{center}
\caption{We depict the heat distribution function $P(Q)$ as function of $Q$ and the 
characteristic function $C(ik)$ as function of $k$. We have set $\beta=1$.}
\label{fig1}
\end{figure}

%%%%%%%%%%%%%%%%%%%%%%%%%%%%%%%%%%%%%%%%%%%%%%%%%%%%%%%%%
%%%%%%%%%%%%%%%%%%%%%%%%%%%%%%%%%%%%%%%%%%%%%%%%%%%%%%%%%
In this paper we have discussed heat fluctuations in equilibrium for an oscillator driven by a single heat reservoir. 
A simple heuristic argument yields the characteristic function $C(k)=\beta^2/(\beta^2-k^2)$ only depending
on the inverse temperature $\beta$. However, the main purpose of the present work is to demonstrate that
this result is also obtained by an explicit representation of the heat reservoir as a collection of independent
oscillators; a representation of a heat bath often used in a quantum mechanical context. We believe this approach
is novel in the context of heat fluctuations. We have discussed 
the problem in terms of complex amplitude variables, the classical counterpart of creation and annihilation
operators for quantum oscillators. The linear coupling to the reservoir is implemented within the rotating
wave approximation, used in quantum optics.  We, moreover, notice that the ohmic approximation,
yielding the usual Langevin description with damping and white noise, here is equivalent to the
quasi particle approximation, known from quantum many body theory. In Fig. \ref{fig1}
we have depicted the characteristic function and the associated heat distribution.
%
%%%%%%%%%%%%%%%%%%%%%%%%%%%%%%%%%%%%%%%%%%%%%%%%%%%%%%%%%
%%%%%%%%%%%%%%%%%%%%%%%%%%%%%%%%%%%%%%%%%%%%%%%%%%%%%%%%%
\section{\label{app}Appendices}
%%%%%%%%%%%%%%%%%%%%%%%%%%%%%%%%%%%%%%%%%%%%%%%%%%%%%%%%%
%%%%%%%%%%%%%%%%%%%%%%%%%%%%%%%%%%%%%%%%%%%%%%%%%%%%%%%%%
%%%%%%%%%%%%%%%%%%%%%%%%
%%%%%%%%%%%%%%%%%%%%%%%%%%%%%%%%%%%%%%%%%%%%%%%%%%%%%%%%%
%%%%%%%%%%%%%%%%%%%%%%%%%%%%%%%%%%%%%%%%%%%%%%%%%%%%%%%%%
\subsection{\label{app1}Multi-oscillator heat bath}
%%%%%%%%%%%%%%%%%%%%%%%%%%%%%%%%%%%%%%%%%%%%%%%%%%%%%%%%%
%%%%%%%%%%%%%%%%%%%%%%%%%%%%%%%%%%%%%%%%%%%%%%%%%%%%%%%%%
The oscillator and the heat bath are described by the Hamiltonian $H_0$ in Eq. (\ref{ham1}) and $H_1$ in 
Eq. (\ref{ham2}), where the coordinates and momenta satisfy the Poisson brackets \cite{Landau59b}
\begin{eqnarray}
&&\{x,p\}=1,
\label{b1}
\\
&&\{x_k,p_p\}=\delta_{pk}.
\label{b2}
\end{eqnarray}
Introducing the complex amplitude variables \cite{Breuer02}
\begin{eqnarray}
&&a=\sqrt\frac{m\omega_0}{2}x+i\frac{1}{\sqrt{2m\omega_0}}p,
\label{b3}
\\
&&b_k=\sqrt\frac{M\Omega_k}{2}x_k+i\frac{1}{\sqrt{2M\Omega_k}}p_k,
\label{b4}
\end{eqnarray}
we have the Poisson brackets
\begin{eqnarray}
&&\{a,a^\ast\}=-i,
\label{b5}
\\
&&\{b_k,b_p^\ast\}=-i\delta_{kp},
\label{b6}
\end{eqnarray}
and the Hamiltonians $H_0$, $H_1$, and $V$ in Eqs. (\ref{ham3}), (\ref{ham4}), and (\ref{ham5}).
Noting that the total Hamiltonian $H=H_0+H_1+V$ is time independent we obtain from the general 
equation of motion $dA/dt=\{A,H\}$:
\begin{eqnarray}
&&i\frac{da}{dt}=\omega_0a+\sum_k\lambda_kb_k,
\label{b7}
\\
&&i\frac{db_k}{dt}=\Omega_kb_k+\lambda_ka.
\label{b8}
\end{eqnarray}
Introducing the Laplace transform \cite{Lebedev72,Mathews73}
\begin{eqnarray}
&&\tilde a(\omega)=\int_0^\infty dt \exp(i\omega t)a(t),
\label{b9}
\\
&&a(t)=\int_{-\infty}^\infty \frac{d\omega}{2\pi} \exp(-i\omega t)\tilde a(\omega),
\label{b10}
\end{eqnarray}
with $\omega$ just above the real axis, i.e., $\omega\rightarrow\omega+i\epsilon$,
the equations of motion take the form
\begin{eqnarray}
&&(\omega-\omega_0)\tilde a(\omega)=ia+\sum_k\lambda_k\tilde b_k(\omega),
\label{b11}
\\
&&(\omega-\Omega_k)\tilde b_k(\omega)=ib_k+\lambda_k\tilde a(\omega),
\label{b12}
\end{eqnarray}
where $a=a(0)$ and $b_k=b_k(0)$ denote the initial values.
Solving for $\tilde a(\omega)$ and  $\tilde b_k(\omega)$ we find 
\begin{eqnarray}
&&(\omega-\omega_0-\Sigma(\omega))\tilde a(\omega)=ia+\sum_k\frac{ib_k\lambda_k}{\omega-\Omega_k},
\label{b13}
\\
&&\sum_p\bigg((\omega-\Omega_p)\delta_{kp}-\Sigma_{kp}(\omega)\bigg)\tilde b_p(\omega)=
ib_k+\frac{ia\lambda_k}{\omega-\omega_0},
\label{b14}
\end{eqnarray}
where we have introduced the self energies
\begin{eqnarray}
&&\Sigma(\omega)=\sum_k\frac{\lambda_k^2}{\omega-\Omega_k},
\label{b15}
\\
&&\Sigma_{kp}(\omega)=\frac{\lambda_k\lambda_p}{\omega-\omega_0}.
\label{b16}
\end{eqnarray}
Using the Plemelj formula $1/(\omega+i\epsilon)=P(1/\omega)-i\pi\delta(\omega)$ we obtain for $\Sigma(\omega)$
\begin{eqnarray}
&&\Sigma(\omega)=\Delta(\omega)-i\Gamma(\omega),
\label{b17}
\\
&&\Delta(\omega)=P\sum_k\frac{\lambda_k^2}{\omega-\Omega_k},
\label{b18}
\\
&&\Gamma(\omega)=\pi\sum_k \lambda_k^2\delta(\omega-\Omega_k).
\label{b19}
\end{eqnarray}
In the weak coupling limit we can make the quasi particle approximation well known in many body
theory \cite{Mahan90} and replace $\omega$ by $\omega_0$, i.e.,
\begin{eqnarray}
&&\Delta(\omega)\sim \Delta(\omega_0)=\Delta,
\label{b20}
\\
&&\Gamma(\omega)\sim\Gamma(\omega_0)=\Gamma.
\label{b21}
\end{eqnarray}
In the present context the weak coupling quasi particle approximation corresponds to
the ohmic approximation  \cite{Ford65,Ford87,Ford88}.
Absorbing the shift $\Delta$ in a renormalisation of $\omega_0$, i.e., defining
$\tilde\omega_0=\omega_0+\Delta$, we obtain for the oscillator amplitude
\begin{eqnarray}
\tilde a(\omega)=\frac{ia}{\omega-\tilde\omega_0+i\Gamma}
+\sum_k\frac{i\lambda_k}{(\omega-\tilde\omega_0+i\Gamma)(\omega-\Omega_k)}b_k,
\label{b22}
\end{eqnarray}
and in time
\begin{eqnarray}
a(t)=ae^{-i\tilde\omega_0t-\Gamma t}+\sum_k\frac{\lambda_k}{\Omega_k-\tilde\omega_0+i\Gamma}
\bigg(e^{-i\Omega_k t}-e^{-i\tilde\omega_0t-\Gamma t}\bigg)b_k.
\label{b23}
\end{eqnarray}
For $t=0$ we have $a(t)=a$; at long times for $t\gg 1/\Gamma$ we obtain Eq. (\ref{amp}), i.e.,
\begin{eqnarray}
a(t)=\sum_k\frac{\lambda_ke^{-i\Omega_k t}}{\Omega_k-\tilde\omega_0+i\Gamma}b_k.
\label{b24}
\end{eqnarray}

Regarding the reservoir modes we obtain solving Eq. (\ref{b14}) for $\tilde b(\omega)$ 
\begin{eqnarray}
\tilde b_p(\omega)=
\frac{ib_p}{\omega-\Omega_p}+\frac{\lambda_p}{(\omega-\Omega_p)(\omega-\tilde\omega_0+i\Gamma)}
\bigg(\sum_k\frac{\lambda_kib_k}{\omega-\Omega_k}+ia\bigg).
\label{b25}
\end{eqnarray}
In time we have
\begin{eqnarray}
b_p(t)=
&&b_pe^{-i\Omega_pt}+a\lambda_p\frac{e^{-i\Omega_pt}-e^{-i\tilde\omega_0 t-\Gamma t}}
{\Omega_p-\tilde\omega_0+i\Gamma}
\nonumber
\\
&&+\sum_k\frac{\lambda_p\lambda_kb_k}{\Omega_p-\Omega_k}\bigg(
\frac{e^{-i\Omega_pt}-e^{-i\tilde\omega_0 t-\Gamma t}}
{\Omega_p-\tilde\omega_0+i\Gamma}- \frac{e^{-i\Omega_kt}-e^{-i\tilde\omega_0 t-\Gamma t}}
{\Omega_k-\tilde\omega_0+i\Gamma}  \bigg).
\label{b26}
\end{eqnarray}
For $t=0$ we have $b_p(t)=b_p$; at long times for $t\gg 1/\Gamma$ we obtain
\begin{eqnarray}
b_p(t)=
&&b_pe^{-i\Omega_pt}+a\lambda_p\frac{e^{-i\Omega_pt}}{\Omega_p-\tilde\omega_0+i\Gamma}
\nonumber
\\
&&+\sum_k\frac{\lambda_p\lambda_kb_k}{\Omega_p-\Omega_k}\bigg(\frac{e^{-i\Omega_pt}}
{\Omega_p-\tilde\omega_0+i\Gamma}- \frac{e^{-i\Omega_kt}}
{\Omega_k-\tilde\omega_0+i\Gamma}  \bigg).
\label{b27}
\end{eqnarray}

In the limit of a large heat reservoir the discrete spectrum of reservoir modes labeled by
the wavenumber $k$ becomes a continuum. Replacing $\sum_k \lambda_k^2\cdots$ by the integral
$\int d\Omega g(\Omega)\cdots$, where the density of states $g(\Omega)$ incorporates
the coupling $\lambda_k$, we obtain
\begin{eqnarray}
&&\Sigma(\omega)=\int d\Omega\frac{g(\Omega)}{\omega-\Omega},
\label{b28}
\\
&&\Delta(\omega)=P\int d\Omega\frac{g(\Omega)}{\omega-\Omega},
\label{b29}
\\
&&\Gamma(\omega)=\pi\int d\Omega g(\Omega)\delta(\omega-\Omega)=\pi g(\omega).
\label{b30}
\end{eqnarray}
The continuum limit thus automatically implies irreversibility and the separation
of time scales. These two assumptions are encoded in the standard classical
Langevin/Fokker Planck approach. The long time expressions for the fields $a(t)$ and $b_p(t)$ 
in Eqs. (\ref{b24}) and (\ref{b27}) are easy to interpret. Regarding $a(t)$ we notice that the dependence 
on the initial value $a$ drops out and $a(t)$ is entirely driven by the heat reservoir thus depending on the initial
heat reservoir characterised by $b_k$; the heat reservoir gives rise to damping and at the 
same time locks $a(t)$ onto $b_k(0)$. Regarding the reservoir amplitudes $b_k(t)$ there is a first order 
"back action" on the k-th mode from the oscillator and an induced second order interaction
between the modes.
%%%%%%%%%%%%%%%%%%%%%%%%%%%%%%%%%%%%%%%%%%%%%%%%%%%%%%%%%
%%%%%%%%%%%%%%%%%%%%%%%%%%%%%%%%%%%%%%%%%%%%%%%%%%%%%%%%%
\subsection{\label{app2}Multi-oscillator heat bath derivation of $P(Q)$}
%%%%%%%%%%%%%%%%%%%%%%%%%%%%%%%%%%%%%%%%%%%%%%%%%%%%%%%%%
%%%%%%%%%%%%%%%%%%%%%%%%%%%%%%%%%%%%%%%%%%%%%%%%%%%%%%%%%
In order to evaluate the distribution $P(Q)$ we must keep track of the energy flow or heat $Q$ between the 
reservoir and the oscillator. In the weak coupling limit we can ignore the energy stored in the interaction
term $V$ given by (\ref{ham5}) and identify the heat $Q$ with the increase of the oscillator energy in
the time span $t$. Consequently, $Q$ is given by Eq. (\ref{heat1}), i.e. $Q(t)=H_0(t)-H_0(0)$, where
$H_0(t)=\omega_0|a(t)|^2$. Inserting the solution $a(t)$ in Eq. (\ref{b24}) we arrive at
\begin{eqnarray}
&&Q(t)=\sum_{kp}b_k^\ast b_pB_{kp}(t),
\label{c1}
\\
&&B_{kp}(t)=A_k^\ast(t)A_p(t)-A_k^\ast(0)A_p(0),
\label{c2}
\\
&& A_k(t)=\sqrt{\omega_0}\frac{\lambda_k e^{-i\Omega_kt}}{\Omega_k-\tilde\omega_0+i\Gamma}.
\label{c3}
\end{eqnarray}

For the characteristic function in Eq. (\ref{char3}) we then obtain
\begin{eqnarray}
C(k)=\int\prod_kdb_kdb_k^\ast e^{-\beta H_1(0)}e^{kQ(t)}/\int\prod_k db_kdb_k^\ast e^{-\beta H_1(0)},
\label{c4}
\end{eqnarray}
where we average over the bath amplitudes $b_k$ at time $t=0$.  Using the identity
\begin{eqnarray}
\int\prod_kdb_kdb_k^\ast e^{-\sum_{pq}A_{pq}b_p^\ast b_q}\propto(\det A)^{-1},
\label{c5}
\end{eqnarray}
and inserting $H_1(0)$ from Eq. (\ref{ham4})
we have
\begin{eqnarray}
C(k)= \frac{\det(\beta\Omega_k\delta_{kp})}{\det(\beta\Omega_k\delta_{kp}-kB_{kp})}.
\label{c6}
\end{eqnarray}
In order to evaluate the determinant we consider the eigenvalue problem
\begin{eqnarray}
\sum_l(\beta\Omega_k\delta_{kl}-kB_{kl})\Phi_l=\mu\Phi_k.
\label{c7}
\end{eqnarray}
Inserting $B_{kl}$ from Eqs. (\ref{c2}) and  (\ref{c3}) we have
\begin{eqnarray}
&&(\beta\Omega_k-\mu)\Phi_k=k(A_k^\ast(t)K-A_k^\ast(0)L),
\label{c8}
\\
&&K=\sum_l A_l(t)\Phi_l,
\label{c9}
\\
&&L=\sum_l A_l(0)\Phi_l.
\label{c10}
\end{eqnarray}
Solving Eq.  (\ref{c8}) for $\Phi_l$ and inserting in Eqs.  (\ref{c9}) and (\ref{c10}) we obtain the linear system
\begin{eqnarray}
&&K=K\sum_l\frac{k|A_l(t)|^2}{\beta\Omega_l-\mu}-
L\sum_l\frac{k A_l(t)A_l^\ast(0)}{\beta\Omega_l-\mu},
\label{c11}
\\
&&L=K\sum_l\frac{k A_l(0)A_l^\ast(t)}{\beta\Omega_l-\mu}-
L\sum_l\frac{k|A_l(0)|^2}{\beta\Omega_l-\mu},
\label{c12}
\end{eqnarray}
implying the determinantal condition
\begin{eqnarray}
&&\bigg(1-\sum_k\frac{k|A_k(t)|^2}{\beta\Omega_k-\mu}\bigg)
\bigg(1+\sum_k\frac{k|A_k(0)|^2}{\beta\Omega_k-\mu}\bigg)
\nonumber
\\
&&+\sum_k\frac{kA_k(t)A_k^\ast(0)}{\beta\Omega_k-\mu}
\sum_p\frac{kA_p(0)A_p^\ast(t)}{\beta\Omega_p-\mu}=0.
\label{c13}
\end{eqnarray}
Further reduction inserting $A_p$ yields the condition
\begin{eqnarray}
k^2\sum_{kp}\Bigg[\frac{\omega_0^2\lambda_k^2\lambda_p^2(1-e^{i(\Omega_k-\Omega_p)t})}
{(\beta\Omega_k-\mu)(\beta\Omega_p-\mu)((\Omega_k-\tilde\omega_0)^2+\Gamma^2)
((\Omega_p-\tilde\omega_0)^2+\Gamma^2)}\Bigg]=1, 
\label{c14}
\end{eqnarray}
determining the eigenvalues $\mu$. In the continuum limit inserting the density of states 
$g(\Omega)$ we have
\begin{eqnarray}
k^2\int d\Omega d\Omega'
\frac{g(\Omega)g(\Omega')\omega_0^2(1-e^{i(\Omega-\Omega')t})}
{(\beta\Omega-\mu)(\beta\Omega'-\mu)((\Omega-\tilde\omega_0)^2+
\Gamma^2)((\Omega'-\tilde\omega_0)^2+\Gamma^2)}=1.
\label{c15}
\end{eqnarray}
Integrating over $\Omega$ and $\Omega'$ to leading order in $\lambda_k$ setting 
$\tilde\omega_0\sim\omega_0$ and introducing $\Gamma=\pi g(\omega_0)$ we obtain
\begin{eqnarray}
k^2g(\omega_0)^2\omega_0^2(\pi/\Gamma)^2\frac{1}{(\beta\omega_0-\mu)^2}=1,
\label{c16}
\end{eqnarray}
or the eigenvalues 
\begin{eqnarray}
\mu_\pm\propto \beta\pm k,
\label{c17}
\end{eqnarray}
yielding
\begin{eqnarray}
C(k)=\frac{\beta^2}{\beta^2-k^2},
\label{c18}
\end{eqnarray}
in agreement with $C(k)$ in Eq. (\ref{char5}).
%\bibliography{/Users/hansfogedby/Library/texmf/bibtex/bib/articles,books}

\end{document}